%% file: main.tex
\documentclass[a4paper]{PoS}

\usepackage{textcomp}
\usepackage{amsfonts} % AMS
\usepackage{amsmath} % AMS
\usepackage{slashed}
\usepackage{pifont}
\usepackage{multirow,lscape}
\usepackage{array}
\usepackage{subcaption}
\usepackage{verbatim}
\usepackage{bm}
\usepackage{comment}
\usepackage{xcolor}
\usepackage{url} %%% URL
\graphicspath{{./fig/}}

\input{macro.tex}

\title{2018 Update on $\epsK$ with lattice QCD inputs}

\ShortTitle{$\epsK$ with lattice QCD inputs}

\author{Jon A.~Bailey, Sunkyu Lee, \speaker{Weonjong Lee} \\
        Lattice Gauge Theory Research Center, CTP, and FPRD, \\
        Department of Physics and Astronomy, \\
        Seoul National University,
        Seoul 08826, South Korea\\
        E-mail: \email{wlee@snu.ac.kr}}

\author{Yong-Chull Jang\\
        Physics Department,
        Brookhaven National Laboratory,
        Upton, NY11973, USA}

\author{Jaehoon Leem\\
        School of Physics,
        Korea Institute for Advanced Study (KIAS),
        Seoul 02455, South Korea}

\author{Sungwoo Park\\
        Los Alamos National Laboratory,
        Theoretical Division T-2,
        Los Alamos, NM87545, USA}

\author{SWME Collaboration}

\abstract{ We present updated results for $\epsK$ determined directly
  from the standard model (SM) with lattice QCD inputs such as $\BK$,
  $\Vcb$, $\Vus$, $\xi_0$, $\xi_2$, $\xi_\text{LD}$, $F_K$, and $m_c$.
  We find that the standard model with exclusive $\Vcb$ and other
  lattice QCD inputs describes only 70\% of the experimental value of
  $|\epsK|$ and does not explain its remaining 30\%, which leads to a
  strong tension in $|\epsK|$ at the $4\sigma$ level between the SM
  theory and experiment.  We also find that this tension disappears
  when we use the inclusive value of $\Vcb$ obtained using the heavy
  quark expansion based on QCD sum rules.  }

\FullConference{The 36th Annual International Symposium on
                Lattice Field Theory - LATTICE2018\\
		22-28 July, 2018\\
		Michigan State University, East Lansing, Michigan, USA.}

\begin{document}

\section{Introduction}
This paper is a brief summary of our previous paper \cite{
  Bailey:2018feb}.
This paper is also an update of our previous papers
\cite{ Jang:2017ieg, Bailey:2015tba, Bailey:2015frw}.

\section{Input parameters: $\Vcb$ and $\xi_0$}
\label{sec:Vcb}

In Table \ref{tab:Vcb}, we present updated results for both exclusive $\Vcb$
and inclusive $\Vcb$.
Recently, HFLAV reported them in Ref.~\cite{ Amhis:2016xyh}.
The results for exclusive $\Vcb$ are obtained using lattice QCD
results for the semileptonic form factors of Refs. \cite{
  Bailey2014:PhysRevD.89.114504, Lattice:2015rga, Detmold:2015aaa}.
Here, we use the combined results (ex-combined) for exclusive $\Vcb$
and the results of the $1S$ scheme for inclusive $\Vcb$ to evaluate
$\epsK$.
For more details on $\Vcb$ and the related caveats, refer to
Ref.~\cite{ Bailey:2018feb}.

\input{tab_Vcb.tex}
%

%
%\section{Input parameter $\xi_0$}
%
The absorptive part of long distance effects on $\epsK$ is parametrized
into $\xi_0$.
\begin{align}
  \xi_0  &= \frac{\Im A_0}{\Re A_0}, \qquad
  \xi_2 = \frac{\Im A_2}{\Re A_2}, \qquad
  \Re \left(\frac{\eps'}{\eps} \right) =
  \frac{\omega}{\sqrt{2} |\eps_K|} (\xi_2 - \xi_0) \,.
  \label{eq:e'/e:xi0}
\end{align}
There are two independent methods to determine $\xi_0$ in lattice QCD:
one is the indirect method and the other is the direct method.
In the indirect method, one can determine $\xi_0$ using
Eq.~\eqref{eq:e'/e:xi0} with lattice QCD input $\xi_2$ and with
experimental results for $\eps'/\eps$, $\epsK$, and $\omega$.
In the direct method, one can determine $\xi_0$ directly using
lattice QCD results for $\Im A_0$ combined with experimental
results for $\Re A_0$.
In Table \ref{tab:xi0+d0}\;(\subref{tab:xi0}), we summarize results for
$\xi_0$ calculated by RBC-UKQCD using the indirect and direct methods.
Here, we use the results of the indirect method for $\xi_0$ to evaluate
$\epsK$.

In Ref.~\cite{ Bai:2015nea}, RBC-UKQCD also reported the S-wave
scattering phase shift for the $I=0$ channel: $\delta_0 =
23.8(49)(12)$, which is different from those of the dispersion
relations \cite{ Colangelo:2001df, GarciaMartin:2011cn} by $\approx 3
\sigma$.
In Ref.~\cite{ Wang:2018Latt}, they have accumulated higher statistics
to obtain $\delta_0 = 19.1(25)(12)$, which is about $5\sigma$
different from those of the dispersion analyses.
They introduce a $\sigma$ operator and make all possible combinations
with the $\sigma$ and $\pi-\pi$ operators.
Then, RBC-UKQCD has obtained $\delta_0 = 32.8(12)(30)$ which is
consistent with those of the dispersion relations.
These results are presented in Table
\ref{tab:xi0+d0}\;(\subref{tab:d0}) and Figure
\ref{tab:xi0+d0}\;(\subref{fig:d0}).

\input{tab_xi0_d0.tex}

\section{Input parameters: Wolfenstein parameters, $\BK$,
  $\xi_\text{LD}$, and others}
In Table \ref{tab:input-WP-eta}\;(\subref{tab:WP}), we summarize the
Wolfenstein parameters on the market.
The CKMfitter and UTfit collaboration provide the Wolfenstein parameters
determined by the global unitarity triangle (UT) fit.
Unfortunately, $\epsK$, $\BK$, and $\Vcb$ are used as inputs to the
global UT fit, which leads to unwanted correlation with $\epsK$.
We want to avoid this correlation, and so take another input set from
the angle-only fit (AOF) suggested in Ref.~\cite{
  Bevan2013:npps241.89}.
The AOF does not use $\epsK$, $\BK$, and $\Vcb$ as input to determine
the UT apex $(\bar{\rho}, \bar{\eta})$.
Here the $\lambda$ parameter is determined from $\Vus$ which is
obtained from the $K_{\ell 2}$ and $K_{\ell 3}$ decays using lattice
QCD results for the form factors and decay constants.
The $A$ parameter is determined from $\Vcb$.

\input{tab_wp_eta.tex}
%

%
%\section{Input parameter $\BK$}
%
%
%
In the FLAG review \cite{Aoki:2016frl}, they present lattice QCD results
for $\BK$ with $N_f=2$, $N_f=2+1$, and $N_f= 2+1+1$.
Here, we use the results for $\BK$ with $N_f=2+1$, which is obtained
by taking a global average over the four data points from BMW 11
\cite{ Durr:2011ap}, Laiho 11 \cite{ Laiho:2011np}, RBC-UKQCD 14
\cite{ Blum:2014tka}, and SWME 15 \cite{ Jang:2015sla}.
In Table \ref{tab:input-BK-other}\;(\subref{tab:BK}), we present the
FLAG 17 result for $\BK$ with $N_f = 2+1$, which is used to evaluate
$\epsK$.

\input{tab_bk_other.tex}

%
%\section{Input parameters: $\xi_\text{LD}$ and others}
%
%
%
The dispersive long distance (LD) effect is defined as
\begin{align}
  \xi_\text{LD} &=  \frac{m^\prime_\text{LD}}{\sqrt{2} \Delta M_K} \,,
  \qquad
  m^\prime_\text{LD}
  = -\Im \left[ \mathcal{P}\sum_{C}
    \frac{\mate{\wbar{K}^0}{H_\text{w}}{C} \mate{C}{H_\text{w}}{K^0}}
         {m_{K^0}-E_{C}}  \right]
  \label{eq:xi-LD}
\end{align}
If the CPT invariance is well respected, the overall contribution
of the $\xi_\text{LD}$ to $\epsK$ is about $\pm 2\%$.
Lattice QCD tools to calculate $\xi_\text{LD}$ are well established in
Refs.~\cite{ Christ2012:PhysRevD.88.014508, Bai:2014cva,
  Christ:2015pwa}.
In addition, there have been a number of attempts to calculate
$\xi_\text{LD}$ on the lattice \cite{ Christ:2015phf, Bai:2016gzv}.
In them, RBC-UKQCD used a pion mass of 329 MeV and a kaon mass of 591
MeV, and so the energy of the 2 pion and 3 pion states are heavier
than the kaon mass.
Hence, the sign of the denominator in Eq.~\ref{eq:xi-LD} is opposite
to that of the physical contribution.
Therefore, this work belongs to the category of exploratory study
rather than to that of precision measurement.
In Ref.~\cite{ Buras2010:PhysLettB.688.309}, they use chiral perturbation
theory to estimate the size of $\xi_\text{LD}$ and claim that
\begin{align}
  \xi_\text{LD} &= -0.4(3) \times \frac{\xi_0}{ \sqrt{2} }
  \label{eq:xiLD:bgi}
\end{align}
where we use the indirect results for $\xi_0$ and its error.
Here, we call this method the BGI estimate for $\xi_\text{LD}$.
In Refs.~\cite{ Christ2012:PhysRevD.88.014508, Christ:2014qwa},
RBC-UKQCD provides another estimate for $\xi_\text{LD}$:
\begin{align}
  \xi_\text{LD} &= (0 \pm 1.6)\%.
  \label{eq:xiLD:rbc}
\end{align}
Here, we call this method the RBC-UKQCD estimate for $\xi_\text{LD}$.
%

%
%\section{Other input parameters}
%
%
%
In Table \ref{tab:input-WP-eta}\;(\subref{tab:eta}), we present
higher order QCD corrections: $\eta_{ij}$ with $i,j = t,c$.
In Table \ref{tab:input-BK-other}\;(\subref{tab:other}), we present
other input parameters needed to evaluate $\epsK$.
Since Lattice 2017, three parameters: $m_t(m_t)$, $m_{K^{0}}$, $F_K$
have been updated.
The $m_t(m_t)$ parameter is the scale-invariant (SI) top quark mass
renormalized in the $\MSb$ scheme.
The pole mass of top quarks comes from Ref.~\cite{Patrignani:2016xqp}:
$  M_t = 173.5 \pm 1.1 \GeV$.
We convert the top quark pole mass into the SI top quark mass using
the four-loop perturbation formula.
For more details, refer to Ref.~\cite{ Bailey:2018feb}.

\section{Results for $\epsK$}
In Fig.~\ref{fig:epsK:cmp:rbc}, we present results for $|\epsK|$
evaluated directly from the standard model (SM) with lattice QCD
inputs given in the previous sections.
In Fig.~\ref{fig:epsK:cmp:rbc}\;(\subref{fig:epsK-ex:rbc}), the blue
curve represents the theoretical evaluation of $|\epsK|$ using the
FLAG-2017 $\BK$, AOF for Wolfenstein parameters, and exclusive $\Vcb$,
and the RBC-UKQCD estimate for $\xi_\text{LD}$.
The red curve in Fig.~\ref{fig:epsK:cmp:rbc} represents the experimental
value of $|\epsK|$.
In Fig.~\ref{fig:epsK:cmp:rbc}\;(\subref{fig:epsK-in:rbc}), the blue
curve represents the same as in
Fig.~\ref{fig:epsK:cmp:rbc}\;(\subref{fig:epsK-ex:rbc})
except for using the inclusive $\Vcb$.

\input{fig_epsK_rbc.tex}

Our results for $|\epsK|$ are summarized in Table \ref{tab:epsK}.
Here, the superscript ${}^\text{SM}$ means that it is obtained
directly from the standard model, the subscript ${}_\text{excl}$
(${}_\text{incl}$) means that it is obtained using exclusive
(inclusive) $\Vcb$, and the superscript ${}^\text{Exp}$ represents the
experimental value.
Results in Table \ref{tab:epsK}\;(\subref{tab:epsK:rbc}) are
obtained using the RBC-UKQCD estimate for $\xi_\text{LD}$ and
those in Table \ref{tab:epsK}\;(\subref{tab:epsK:bgi}) are obtained
using the BGI estimate for $\xi_\text{LD}$.
In Table \ref{tab:epsK}\;(\subref{tab:epsK:rbc}), we find that the
theoretical evaluation of $|\epsK|$ with lattice QCD inputs (with
exclusive $\Vcb$) $|\epsK|^\text{SM}_\text{excl}$ has $4.2\sigma$
tension with the experimental result $|\epsK|^\text{Exp}$, while there
is no tension with inclusive $\Vcb$ (heavy quark expansion with QCD
sum rules).

\input{tab_epsK.tex}

In Fig.~\ref{fig:depsK:sum:rbc:his}\;(\subref{fig:depsK:rbc:his}), we
plot the $\Delta \epsK \equiv |\epsK|^\text{Exp} -
|\epsK|^\text{SM}_\text{excl}$ in units of $\sigma$ (the total error)
as a function of time starting from 2012.
In 2012, $\Delta \epsK$ was $2.5\sigma$, but now it is $4.2\sigma$.
In Fig.~\ref{fig:depsK:sum:rbc:his}\;(\subref{fig:depsK+sigma:rbc:his}),
we plot the history of the average $\Delta \epsK$ and the error
$\sigma_{\Delta \epsK}$.
We find that the average has increased with some fluctuations by 27\%
during the period of 2012-2018, and its error has decreased
monotonically by 25\% in the same period.

\input{fig_epsK_history_rbc.tex}

In Table \ref{tab:err-bud+his-DepsK}\;(\subref{tab:err-budget:rbc}),
we present the error budget for $|\epsK|^\text{SM}_\text{excl}$.
Here, we find that the largest error comes from $\Vcb$.
Hence, it is essential to reduce the error in $\Vcb$ significantly.
%

\input{tab_err_bud_rbc.tex}
In Table \ref{tab:err-bud+his-DepsK}\;(\subref{tab:DepsK}), we present
how the values of $\Delta\epsK$ have changed from 2015 \cite{
  Bailey:2015tba} to 2018 \cite{ Bailey:2018feb}.
Here, we find that the positive shift of $\Delta \epsK$ is about the
same for the inclusive and exclusive $\Vcb$.
This reflects the changes in other parameters since 2015.

\acknowledgments
We thank Shoji Hashimoto and Takashi Kaneko for helpful discussion on
$\Vcb$.
The research of W.~Lee is supported by the Creative Research
Initiatives Program (No.~2017013332) of the NRF grant funded by the
Korean government (MEST).
~J.A.B. is supported by the Basic Science Research Program of the
National Research Foundation of Korea (NRF) funded by the Ministry of
Education (No.~2015024974).
W.~Lee would like to acknowledge the support from the KISTI
supercomputing center through the strategic support program for the
supercomputing application research (No.~KSC-2016-C3-0072).
Computations were carried out on the DAVID GPU clusters at Seoul
National University.

%------------
% references
%------------
\bibliography{refs}

%----------
% all done
%----------

\end{document}

%% file: macro.tex
\providecommand{\wbar}[1]{\overline#1}

\providecommand{\mate}[3]{\langle#1\lvert#2\rvert#3\rangle}

\renewcommand{\Re}{\mathrm{Re}\,}
\renewcommand{\Im}{\mathrm{Im}\,}

\providecommand{\GeV}{\;\mathrm{GeV}}

\definecolor{HLBlue}{HTML}{6599FF}
\definecolor{HLOrange}{HTML}{FF6600}

\newcommand{\MSb}{\overline{\mathrm{MS}}}

\newcommand{\BK}{\hat{B}_{K}}
\newcommand{\Vcb}{|V_{cb}|}
\newcommand{\Vub}{|V_{ub}|}
\newcommand{\Vus}{|V_{us}|}

\newcommand{\eps}{\varepsilon}

\newcommand{\epsK}{\varepsilon_{K}}

%% file: tab_Vcb.tex
\begin{table}[t!]
  \begin{subtable}{0.44\linewidth}
    \center
    \vspace*{-7mm}
    \resizebox{0.9\textwidth}{!}{
      \begin{tabular}{l l l}
        \hline\hline
        channel & value & Ref. \\ \hline
        $B\to D^* \ell \bar{\nu}$
        & $39.05(47)(58)$ &
        \cite{Amhis:2016xyh, Bailey2014:PhysRevD.89.114504} \\
        $B\to D \ell \bar{\nu}$
        & $39.18(94)(36)$ &
        \cite{Amhis:2016xyh, Lattice:2015rga}  \\
        $|V_{ub}|/|V_{cb}|$
        & $0.080(4)(4)$   &
        \cite{Amhis:2016xyh, Detmold:2015aaa} \\ \hline
        ex-combined
        & $39.13(59)$     & \cite{Amhis:2016xyh}
        \\ \hline\hline
      \end{tabular}
    } %%% \resizebox
    \caption{Results for exclusive $|V_{cb}|$.}
    \label{tab:ex-Vcb}
    \vspace{7mm}
    \resizebox{0.9\textwidth}{!}{
      \begin{tabular}{l l c}
        \hline\hline
        channel & value & Ref. \\ \hline
        kinetic scheme & $42.19(78)$ & \cite{Amhis:2016xyh}
        \\
        1S scheme      & $41.98(45)$ & \cite{Amhis:2016xyh}
        \\ \hline
        in-combined    & $42.03(39)$ & \cite{Bailey:2018feb}
        \\ \hline\hline
      \end{tabular}
    } %%% \resizebox
    \caption{Results for inclusive $|V_{cb}|$.}
    \label{tab:in-Vcb}
  \end{subtable}
  \hfill
  \begin{subfigure}{0.54\linewidth}
    \vspace*{-7mm}
    \centering
    \includegraphics[width=\textwidth]{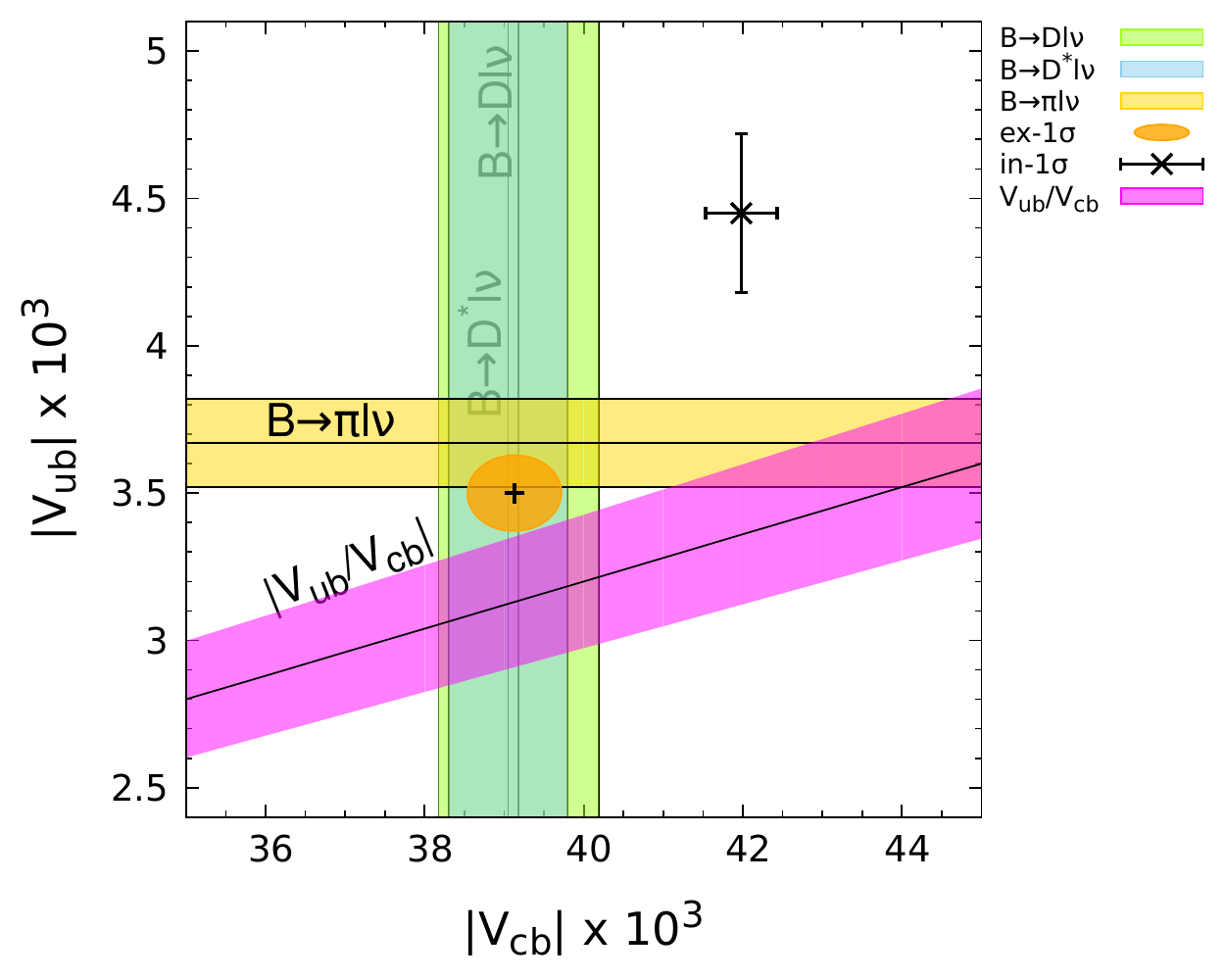}
    \caption{$|V_{cb}|$ versus $|V_{ub}|$.}
    \label{fig:Vcb-Vub}
  \end{subfigure}
  %%%
  \caption{ $\Vcb$: (\subref{tab:ex-Vcb}) exclusive $\Vcb$,
    (\subref{tab:in-Vcb}) inclusive $\Vcb$, and (\subref{fig:Vcb-Vub})
    $\Vcb$ versus $\Vub$. }
  \label{tab:Vcb}
\end{table}

%% file: tab_xi0_d0.tex
\begin{table}[t!]
  \begin{subtable}{0.44\linewidth}
    \renewcommand{\arraystretch}{1.2}
    \vspace*{-7mm}
    \begin{center}
      \resizebox{0.8\linewidth}{!}{
        \begin{tabular}{lcc}
          \hline\hline
          method & $\xi_0$ & Ref. \\ \hline
          indirect & $-1.63(19) \times 10^{-4}$ & \cite{Blum:2015ywa}
          \\
          direct  & $-0.57(49) \times 10^{-4}$  & \cite{Bai:2015nea}
          \\ \hline\hline
        \end{tabular}
      } % resizebox
      \caption{$\xi_0$ }
      \label{tab:xi0}
    \end{center}
    \resizebox{1.0\linewidth}{!}{
      \begin{tabular}{llr}
        \hline\hline
        Collaboration & $\delta_0$ & Ref. \\ \hline
        RBC-UKQCD-15  & $23.8(49)(12){}^{\circ}$ & \cite{Bai:2015nea}
        \\
        RBC-UKQCD-18  & $32.8(12)(30){}^{\circ}$ & \cite{Wang:2018Latt}
        \\
        KPY-2011 & 39.1(6)${}^{\circ}$ & \cite{GarciaMartin:2011cn}
        \\
        CGL-2001 & 39.2(15)${}^{\circ}$ & \cite{Colangelo:2001df,
          Colangelo2016:MITP}
        \\ \hline\hline
      \end{tabular}
    } %%% \resizebox
    \caption{Results for $\delta_0$ }
    \label{tab:d0}
  \end{subtable}
  \hfill
  \begin{subfigure}{0.56\linewidth}
    \vspace*{-7mm}
%%    \hspace{2mm}
    \includegraphics[width=\linewidth]{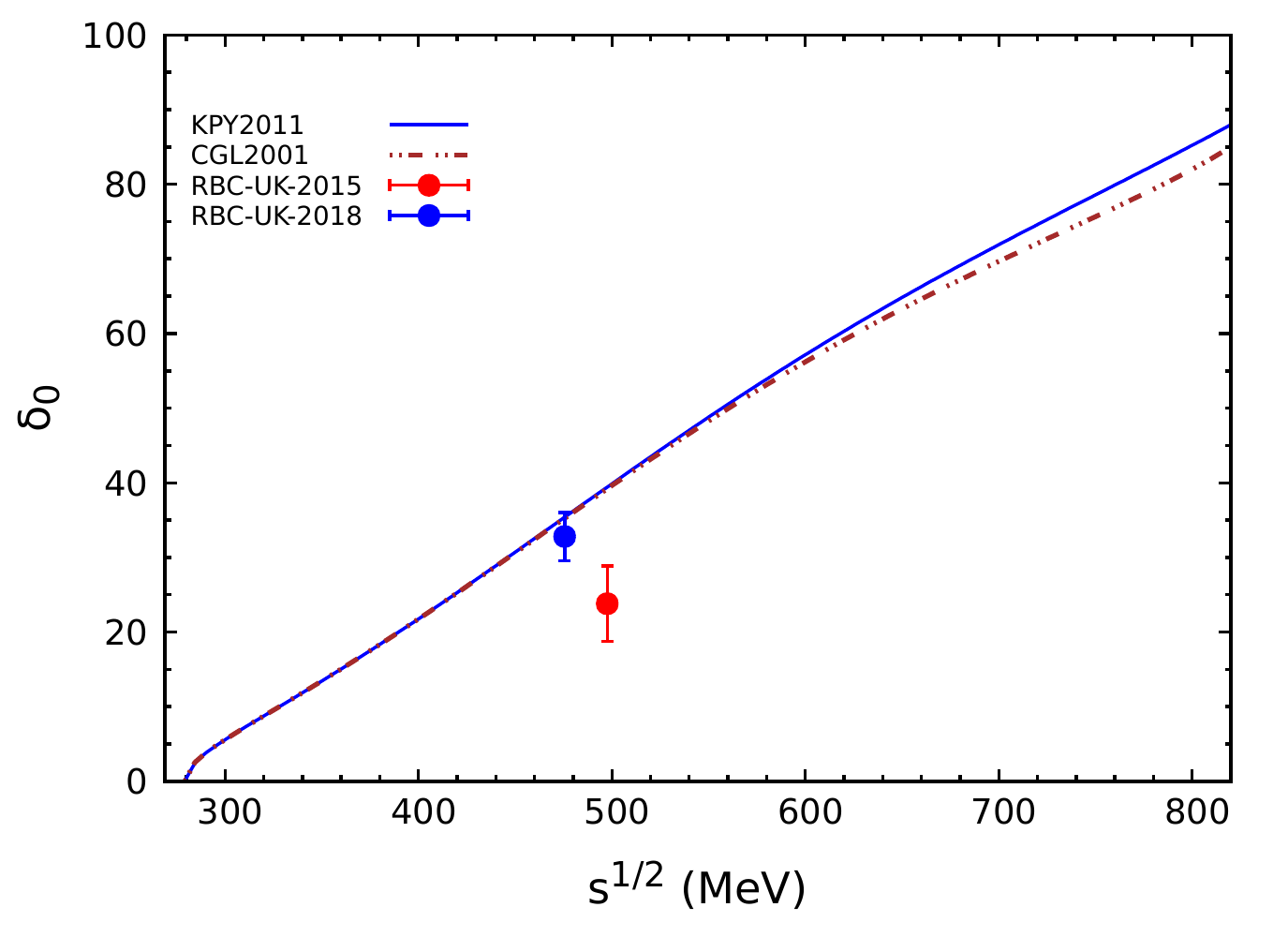}
    \caption{$\delta_0$}
    \label{fig:d0}
  \end{subfigure}
  \caption{The absorptive long distance effect $\xi_0$ and S-wave
    $I=0$ scattering phase shift $\delta_0$. }
  \label{tab:xi0+d0}
\end{table}

%% file: tab_wp_eta.tex
\begin{table}[t!]
%  \footnotesize
%%%  \renewcommand{\arraystretch}{1.2}
%%%  \renewcommand{\subfigcapskip}{0.55em}
%
  \begin{subtable}{0.60\linewidth}
    \renewcommand{\arraystretch}{1.0}
    \vspace*{-3mm}
    \resizebox{1.0\linewidth}{!}{
      \begin{tabular}{ c | l c | l c | l c }
        \hline\hline
        WP
        & \multicolumn{2}{c|}{CKMfitter}
        & \multicolumn{2}{c|}{UTfit}
        & \multicolumn{2}{c}{AOF} \\ \hline
        $\lambda$
        & $0.22509(29)$ & \cite{Charles:2004jd}
        & $0.22497(69)$ & \cite{Bona:2006ah}
        & $0.2248(6)$   & \cite{Patrignani:2016xqp}
        \\ \hline
        $\bar{\rho}$
        & $0.1598(76)$ & \cite{Charles:2004jd}
        & $0.153(13)$  & \cite{Bona:2006ah}
        & $0.146(22)$  & \cite{Martinelli:2017}
        \\ \hline
        $\bar{\eta}$
        & $0.3499(63)$ & \cite{Charles:2004jd}
        & $0.343(11)$  & \cite{Bona:2006ah}
        & $0.333(16)$  & \cite{Martinelli:2017}
        \\ \hline\hline
      \end{tabular}
    } % resizebox
    \caption{Wolfenstein parameters}
    \label{tab:WP}
  \end{subtable} %%% \subtable
  \hfill
  \begin{subtable}{0.26\linewidth}
    \renewcommand{\arraystretch}{1.0}
    \vspace*{-3mm}
    \resizebox{1.0\linewidth}{!}{
      \begin{tabular}[b]{ c l c }
        \hline\hline
        Input & Value & Ref.
        \\ \hline
        $\eta_{cc}$ & $1.72(27)$   & \cite{Bailey:2015tba}
        \\
        $\eta_{tt}$ & $0.5765(65)$ & \cite{Buras2008:PhysRevD.78.033005}
        \\
        $\eta_{ct}$ & $0.496(47)$  & \cite{Brod2010:prd.82.094026}
        \\ \hline\hline
      \end{tabular}
    } % resizebox
    \caption{$\eta_{ij}$}
    \label{tab:eta}
  \end{subtable} %%% \subtable
  \caption{ (\subref{tab:WP}) Wolfenstein parameters and
    (\subref{tab:eta}) QCD corrections: $\eta_{ij}$ with $i,j = c,t$.}
  \label{tab:input-WP-eta}
\end{table}

%% file: tab_bk_other.tex
\begin{table}[t!]
%  \footnotesize
%%%  \renewcommand{\arraystretch}{1.2}
%%%  \renewcommand{\subfigcapskip}{0.55em}
%
  \begin{subtable}{0.40\linewidth}
    \renewcommand{\arraystretch}{1.45}
    \vspace*{-5mm}
    \resizebox{1.0\linewidth}{!}{
      \begin{tabular}{ l  c  l }
        \hline\hline
        Collaboration & Ref. & $\BK$  \\ \hline
        SWME 15       & \cite{Jang:2015sla} & $0.735(5)(36)$     \\
        RBC/UKQCD 14  & \cite{Blum:2014tka} & $0.7499(24)(150)$  \\
        Laiho 11      & \cite{Laiho:2011np} & $0.7628(38)(205)$  \\
        BMW 11        & \cite{Durr:2011ap}  & $0.7727(81)(84)$  \\ \hline
        FLAG 17       & \cite{Aoki:2016frl} & $0.7625(97)$
        \\ \hline\hline
      \end{tabular}
    } % resizebox
    \caption{$\BK$}
    \label{tab:BK}
  \end{subtable} %%% \subtable
  \hfill
  \begin{subtable}{0.40\linewidth}
    \vspace*{-5mm}
    \resizebox{1.0\linewidth}{!}{
      \begin{tabular}{ c l c }
        \hline\hline
        Input & Value & Ref. \\ \hline
        $G_{F}$
        & $1.1663787(6) \times 10^{-5}$ GeV$^{-2}$
        &\cite{Patrignani:2016xqp} \\ \hline
        $M_{W}$
        & $80.385(15)$ GeV
        &\cite{Patrignani:2016xqp} \\ \hline
        $m_{c}(m_{c})$
        & $1.2733(76)$ GeV
        &\cite{Chakraborty:2014aca} \\ \hline
        $m_{t}(m_{t})$
        & $163.65(105)(17)$ GeV
        &\cite{Bailey:2018feb} \\ \hline
        $\theta$
        & $43.52(5)^{\circ}$
        &\cite{Patrignani:2016xqp} \\ \hline
        $m_{K^{0}}$
        & $497.611(13)$ MeV
        &\cite{Patrignani:2016xqp} \\ \hline
        $\Delta M_{K}$
        & $3.484(6) \times 10^{-12}$ MeV
        &\cite{Patrignani:2016xqp} \\ \hline
        $F_K$
        & $155.6(4)$ MeV
        &\cite{Patrignani:2016xqp}
        \\ \hline\hline
      \end{tabular}
    } % resizebox
    \caption{Other parameters}
    \label{tab:other}
  \end{subtable} %%% \subtable
  \caption{ (\subref{tab:BK}) Results for $\BK$ and
    (\subref{tab:other}) other input parameters.}
  \label{tab:input-BK-other}
\end{table}

%% file: fig_epsK_rbc.tex
\begin{figure}[t!]
  \begin{subfigure}{0.47\linewidth}
    \vspace*{-7mm}
    \includegraphics[width=1.0\linewidth]
                    {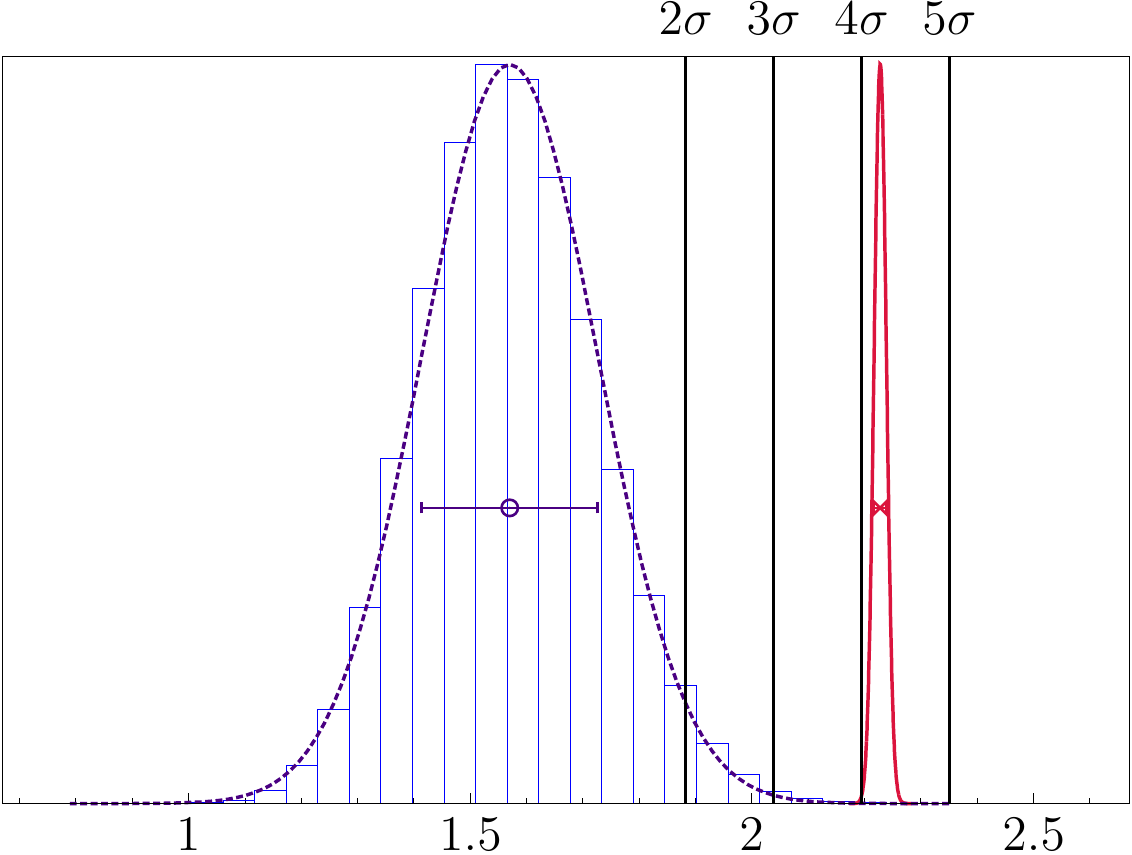}
    \caption{Exclusive $\Vcb$}
    \label{fig:epsK-ex:rbc}
  \end{subfigure}
  \hfill
  \begin{subfigure}{0.47\linewidth}
    \vspace*{-7mm}
    \includegraphics[width=1.0\linewidth]
                    {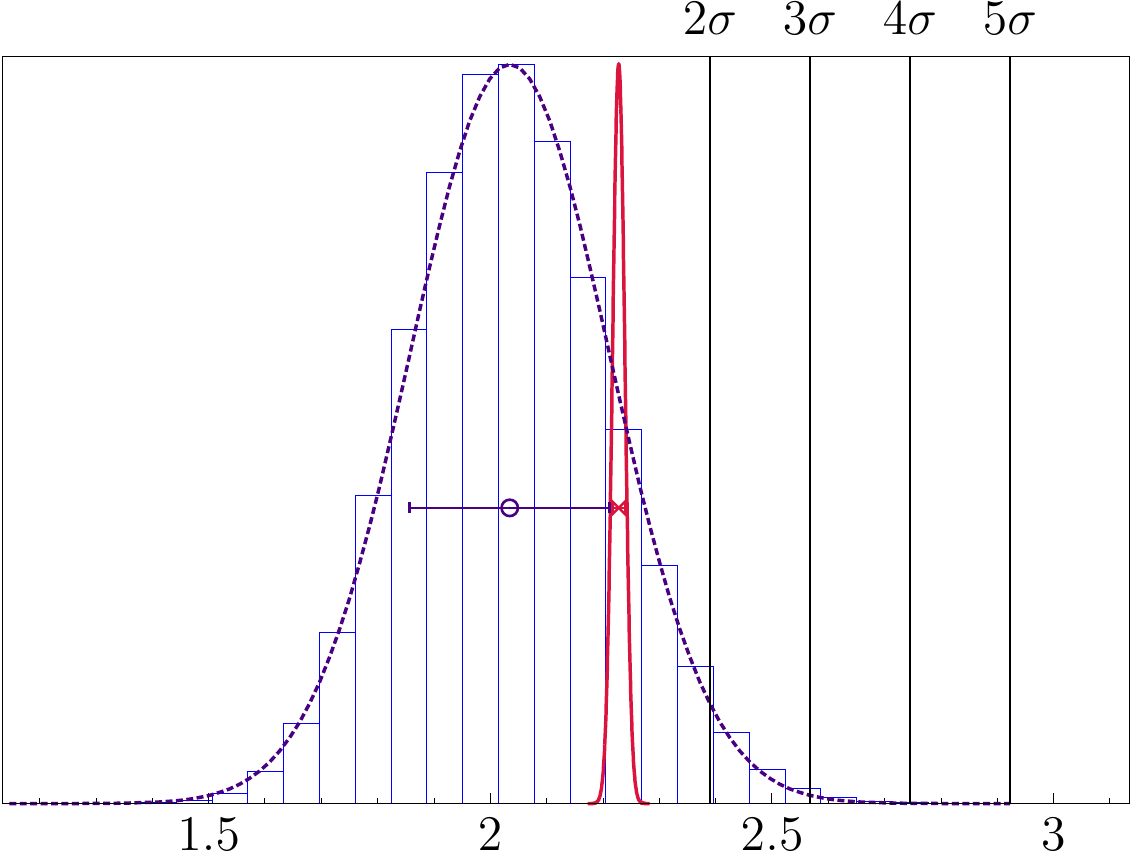}
    \caption{Inclusive $\Vcb$}
    \label{fig:epsK-in:rbc}
  \end{subfigure}
  \caption{$|\epsK|$ with (\subref{fig:epsK-ex:rbc}) exclusive $\Vcb$
    (left) and (\subref{fig:epsK-in:rbc}) inclusive $\Vcb$ (right) in
    units of $1.0\times 10^{-3}$.}
  \label{fig:epsK:cmp:rbc}
\end{figure}

%% file: tab_epsK.tex
\begin{table}[tbhp]
%  \footnotesize
%%%  \renewcommand{\arraystretch}{1.2}
%%%  \renewcommand{\subfigcapskip}{0.55em}
%
  \begin{subtable}{0.50\linewidth}
    \renewcommand{\arraystretch}{1.0}
    \resizebox{1.0\linewidth}{!}{
      \begin{tabular}{lll}
        \hline\hline
        parameter & method & value \\ \hline
        $|\epsK|^\text{SM}_\text{excl}$ & exclusive $\Vcb$ & $1.570 \pm 0.156$
        \\
        $|\epsK|^\text{SM}_\text{incl}$ & inclusive $\Vcb$ & $2.035 \pm 0.178$
        \\
        $|\epsK|^\text{Exp}$            & experiment       & $2.228 \pm 0.011$
        \\ \hline\hline
      \end{tabular}
    } % resizebox
    \caption{RBC-UKQCD estimate for $\xi_\text{LD}$}
    \label{tab:epsK:rbc}
  \end{subtable} %%% \subtable
  \hfill
  \begin{subtable}{0.50\linewidth}
    \renewcommand{\arraystretch}{1.0}
    \resizebox{1.0\linewidth}{!}{
      \begin{tabular}{lll}
        \hline\hline
        parameter & method & value \\ \hline
        $|\epsK|^\text{SM}_\text{excl}$ & exclusive $\Vcb$ & $1.615 \pm 0.158$
        \\
        $|\epsK|^\text{SM}_\text{incl}$ & inclusive $\Vcb$ & $2.079 \pm 0.178$
        \\
        $|\epsK|^\text{Exp}$            & experiment       & $2.228 \pm 0.011$
        \\ \hline\hline
      \end{tabular}
    } % resizebox
    \caption{BGI estimate for $\xi_\text{LD}$}
    \label{tab:epsK:bgi}
  \end{subtable} %%% \subtable
  \caption{ $|\epsK|$ in units of $1.0\times 10^{-3}$.}
  \label{tab:epsK}
\end{table}

%% file: fig_epsK_history_rbc.tex
\begin{figure}[htbp]
  \begin{subfigure}{0.5\linewidth}
    \vspace*{-7mm}
    \includegraphics[width=\linewidth]{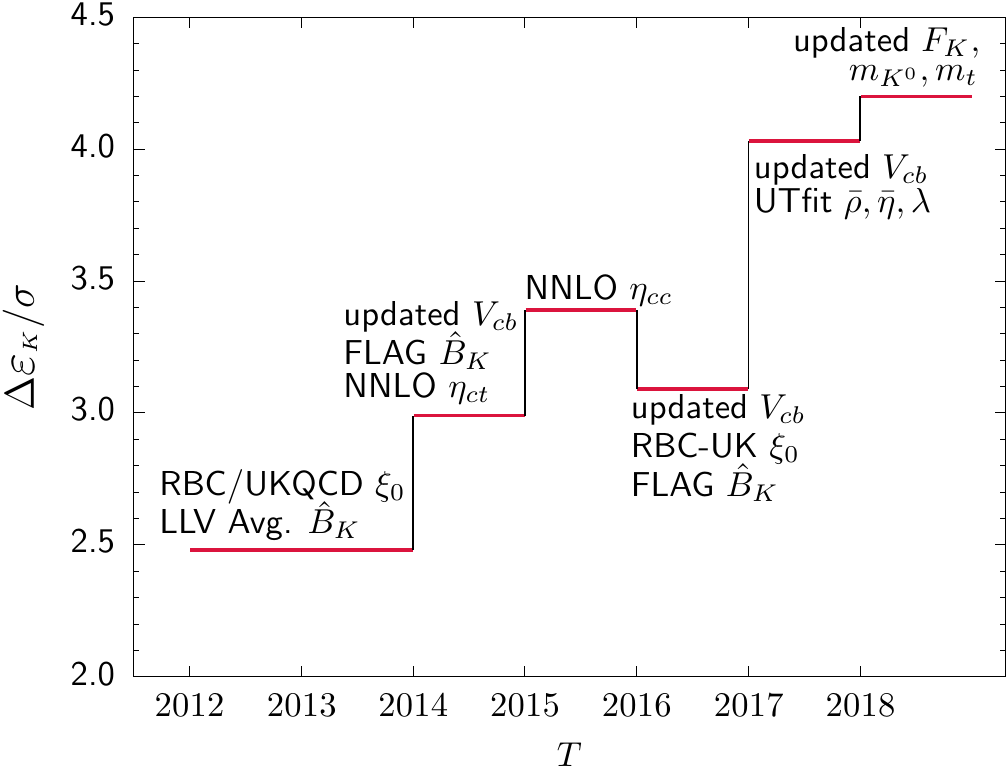}
    \caption{Time evolution of $\Delta \epsK/\sigma$}
    \label{fig:depsK:rbc:his}
  \end{subfigure}
  \hfill
  \begin{subfigure}{0.475\linewidth}
    \vspace*{-7mm}
    \includegraphics[width=\linewidth]
                    {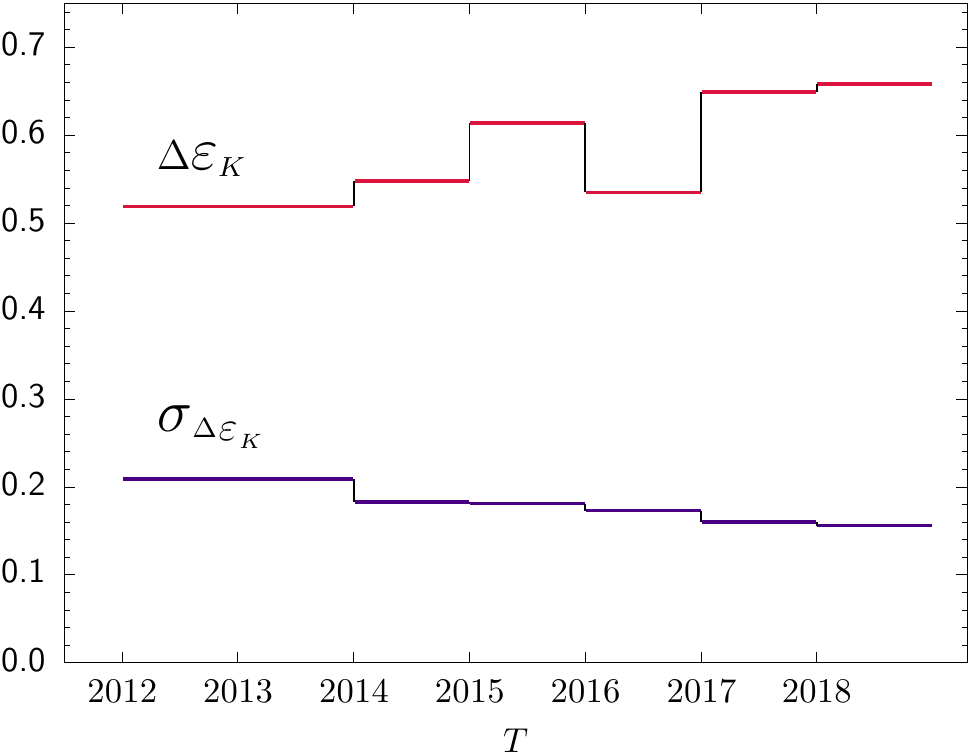}
    \caption{Time evolution of the average and error of $\Delta\epsK$}
    \label{fig:depsK+sigma:rbc:his}
  \end{subfigure}
  \caption{ Time history of (\subref{fig:depsK:rbc:his})
    $\Delta\epsK/\sigma$, and (\subref{fig:depsK+sigma:rbc:his})
    $\Delta\epsK$ and $\sigma_{\Delta\epsK}$. }
  \label{fig:depsK:sum:rbc:his}
\end{figure}

%% file: tab_err_bud_rbc.tex
%--------------
% Error budget
%--------------
\begin{table}[tb!]
  \begin{subtable}{0.35\linewidth}
    \renewcommand{\arraystretch}{1.0}
    \resizebox{1.0\linewidth}{!}{
      \begin{tabular}{ccc}
        \hline\hline
        source          & error (\%)  & memo \\
        \hline
        $\Vcb$          & 31.4        & ex-combined \\
        $\bar{\eta}$    & 26.8        & AOF \\
        $\eta_{ct}$     & 21.5        & $c-t$ Box \\
        $\eta_{cc}$     &  9.1        & $c-c$ Box \\
        $\vdots$        & $\vdots$    & $\vdots$ \\
        \hline\hline
      \end{tabular}
    } %%% end of resizebox
    \caption{Error budget for $|\epsK|^\text{SM}_\text{excl}$}
    \label{tab:err-budget:rbc}
  \end{subtable}
  \hfill
  \begin{subtable}{0.50\linewidth}
    \renewcommand{\arraystretch}{1.2}
    \resizebox{1.0\linewidth}{!}{
      \begin{tabular}{c @{\qquad} c c}
        \hline\hline
        year & Inclusive $\Vcb$ & Exclusive $\Vcb$ \\ \hline
        2015 & $0.33\sigma$     & $3.4\sigma$ \\
        2018 & $1.1\sigma$      & $4.2\sigma$ \\
        \hline\hline
      \end{tabular}
    } %%% end of resizebox
    \caption{Results for $\Delta\epsK$.}
    \label{tab:DepsK}
  \end{subtable}
  \caption{Error budget for $|\epsK|^\text{SM}_\text{excl}$
    and history of $\Delta\epsK$}
  \label{tab:err-bud+his-DepsK}
\end{table}

%% file: main.bbl
\providecommand{\href}[2]{#2}\begingroup\raggedright\begin{thebibliography}{10}

\bibitem{Bailey:2018feb}
J.~A. Bailey, S.~Lee, W.~Lee, J.~Leem, and S.~Park
  \href{http://xxx.lanl.gov/abs/1808.09657}{{\tt 1808.09657}}.

\bibitem{Jang:2017ieg}
Y.-C. Jang, W.~Lee, S.~Lee, and J.~Leem {\em EPJ Web Conf.} {\bf 175} (2018)
  14015, [\href{http://xxx.lanl.gov/abs/1710.06614}{{\tt 1710.06614}}].

\bibitem{Bailey:2015tba}
J.~A. Bailey, Y.-C. Jang, W.~Lee, and S.~Park {\em Phys. Rev.} {\bf D92}
  (2015), no.~3 034510, [\href{http://xxx.lanl.gov/abs/1503.05388}{{\tt
  1503.05388}}].

\bibitem{Bailey:2015frw}
J.~A. Bailey, Y.-C. Jang, W.~Lee, and S.~Park {\em PoS} {\bf LATTICE2015}
  (2015) 348, [\href{http://xxx.lanl.gov/abs/1511.00969}{{\tt 1511.00969}}].

\bibitem{Amhis:2016xyh}
Y.~Amhis {\em et~al.} {\em Eur. Phys. J.} {\bf C77} (2017), no.~12 895,
  [\href{http://xxx.lanl.gov/abs/1612.07233}{{\tt 1612.07233}}].

\bibitem{Bailey2014:PhysRevD.89.114504}
J.~A. Bailey, A.~Bazavov, C.~Bernard, {\em et~al.} {\em Phys.Rev.} {\bf D89}
  (2014) 114504, [\href{http://xxx.lanl.gov/abs/1403.0635}{{\tt 1403.0635}}].

\bibitem{Lattice:2015rga}
J.~A. Bailey {\em et~al.} {\em Phys. Rev.} {\bf D92} (2015), no.~3 034506,
  [\href{http://xxx.lanl.gov/abs/1503.07237}{{\tt 1503.07237}}].

\bibitem{Detmold:2015aaa}
W.~Detmold, C.~Lehner, and S.~Meinel {\em Phys. Rev.} {\bf D92} (2015), no.~3
  034503, [\href{http://xxx.lanl.gov/abs/1503.01421}{{\tt 1503.01421}}].

\bibitem{Bai:2015nea}
Z.~Bai {\em et~al.} {\em Phys. Rev. Lett.} {\bf 115} (2015), no.~21 212001,
  [\href{http://xxx.lanl.gov/abs/1505.07863}{{\tt 1505.07863}}].

\bibitem{Colangelo:2001df}
G.~Colangelo, J.~Gasser, and H.~Leutwyler {\em Nucl. Phys.} {\bf B603} (2001)
  125--179, [\href{http://xxx.lanl.gov/abs/hep-ph/0103088}{{\tt
  hep-ph/0103088}}].

\bibitem{GarciaMartin:2011cn}
R.~Garcia-Martin, R.~Kaminski, J.~R. Pelaez, J.~Ruiz~de Elvira, and F.~J.
  Yndurain {\em Phys. Rev.} {\bf D83} (2011) 074004,
  [\href{http://xxx.lanl.gov/abs/1102.2183}{{\tt 1102.2183}}].

\bibitem{Wang:2018Latt}
T.~Wang.
  {\url{https://indico.fnal.gov/event/15949/session/3/contribution/150/material/slides/0.pdf}
  }.

\bibitem{Blum:2015ywa}
T.~Blum {\em et~al.} {\em Phys. Rev.} {\bf D91} (2015), no.~7 074502,
  [\href{http://xxx.lanl.gov/abs/1502.00263}{{\tt 1502.00263}}].

\bibitem{Colangelo2016:MITP}
\url{https://indico.mitp.uni-mainz.de/event/48/contribution/5/material/slides/0.pdf}.

\bibitem{Bevan2013:npps241.89}
A.~Bevan, M.~Bona, M.~Ciuchini, D.~Derkach, E.~Franco, {\em et~al.} {\em
  Nucl.Phys.Proc.Suppl.} {\bf 241-242} (2013) 89--94.

\bibitem{Charles:2004jd}
J.~Charles {\em et~al.} {\em Eur.Phys.J.} {\bf C41} (2005) 1--131,
  [\href{http://xxx.lanl.gov/abs/hep-ph/0406184}{{\tt hep-ph/0406184}}].
  updated results and plots available at: \url{http://ckmfitter.in2p3.fr}.

\bibitem{Bona:2006ah}
M.~Bona {\em et~al.} {\em JHEP} {\bf 10} (2006) 081,
  [\href{http://xxx.lanl.gov/abs/hep-ph/0606167}{{\tt hep-ph/0606167}}].
  {Standard Model fit results: Summer 2016 (ICHEP 2016):
  \url{http://www.utfit.org}}.

\bibitem{Patrignani:2016xqp}
C.~Patrignani {\em et~al.} {\em Chin. Phys.} {\bf C40} (2016), no.~10 100001.
  {\url{https://pdg.lbl.gov/} }.

\bibitem{Martinelli:2017}
G.~Martinelli {\em et~al.} \url{http://www.utfit.org/UTfit/}, 2017.

\bibitem{Buras2008:PhysRevD.78.033005}
A.~J. Buras and D.~Guadagnoli {\em Phys.Rev.} {\bf D78} (2008) 033005,
  [\href{http://xxx.lanl.gov/abs/0805.3887}{{\tt 0805.3887}}].

\bibitem{Brod2010:prd.82.094026}
J.~Brod and M.~Gorbahn {\em Phys.Rev.} {\bf D82} (2010) 094026,
  [\href{http://xxx.lanl.gov/abs/1007.0684}{{\tt 1007.0684}}].

\bibitem{Aoki:2016frl}
S.~Aoki {\em et~al.} {\em Eur. Phys. J.} {\bf C77} (2017), no.~2 112,
  [\href{http://xxx.lanl.gov/abs/1607.00299}{{\tt 1607.00299}}].

\bibitem{Durr:2011ap}
S.~Durr {\em et~al.} {\em Phys. Lett.} {\bf B705} (2011) 477--481,
  [\href{http://xxx.lanl.gov/abs/1106.3230}{{\tt 1106.3230}}].

\bibitem{Laiho:2011np}
J.~Laiho and R.~S. Van~de Water {\em PoS} {\bf LATTICE2011} (2011) 293,
  [\href{http://xxx.lanl.gov/abs/1112.4861}{{\tt 1112.4861}}].

\bibitem{Blum:2014tka}
T.~Blum {\em et~al.} {\em Phys. Rev.} {\bf D93} (2016), no.~7 074505,
  [\href{http://xxx.lanl.gov/abs/1411.7017}{{\tt 1411.7017}}].

\bibitem{Jang:2015sla}
B.~J. Choi {\em et~al.} {\em Phys. Rev.} {\bf D93} (2016), no.~1 014511,
  [\href{http://xxx.lanl.gov/abs/1509.00592}{{\tt 1509.00592}}].

\bibitem{Chakraborty:2014aca}
B.~Chakraborty, C.~T.~H. Davies, B.~Galloway, P.~Knecht, J.~Koponen, G.~C.
  Donald, R.~J. Dowdall, G.~P. Lepage, and C.~McNeile {\em Phys. Rev.} {\bf
  D91} (2015), no.~5 054508, [\href{http://xxx.lanl.gov/abs/1408.4169}{{\tt
  1408.4169}}].

\bibitem{Christ2012:PhysRevD.88.014508}
N.~Christ, T.~Izubuchi, C.~Sachrajda, A.~Soni, and J.~Yu {\em Phys.Rev.} {\bf
  D88} (2013), no.~1 014508, [\href{http://xxx.lanl.gov/abs/1212.5931}{{\tt
  1212.5931}}].

\bibitem{Bai:2014cva}
Z.~Bai, N.~Christ, T.~Izubuchi, C.~Sachrajda, A.~Soni, {\em et~al.} {\em
  Phys.Rev.Lett.} {\bf 113} (2014), no.~11 112003,
  [\href{http://xxx.lanl.gov/abs/1406.0916}{{\tt 1406.0916}}].

\bibitem{Christ:2015pwa}
N.~H. Christ, X.~Feng, G.~Martinelli, and C.~T. Sachrajda {\em Phys. Rev.} {\bf
  D91} (2015), no.~11 114510, [\href{http://xxx.lanl.gov/abs/1504.01170}{{\tt
  1504.01170}}].

\bibitem{Christ:2015phf}
N.~H. Christ and Z.~Bai {\em PoS} {\bf LATTICE2015} (2016) 342.

\bibitem{Bai:2016gzv}
Z.~Bai {\em PoS} {\bf LATTICE2016} (2017) 309,
  [\href{http://xxx.lanl.gov/abs/1611.06601}{{\tt 1611.06601}}].

\bibitem{Buras2010:PhysLettB.688.309}
A.~J. Buras, D.~Guadagnoli, and G.~Isidori {\em Phys.Lett.} {\bf B688} (2010)
  309--313, [\href{http://xxx.lanl.gov/abs/1002.3612}{{\tt 1002.3612}}].

\bibitem{Christ:2014qwa}
N.~Christ, T.~Izubuchi, C.~T. Sachrajda, A.~Soni, and J.~Yu {\em PoS} {\bf
  LATTICE2013} (2014) 397, [\href{http://xxx.lanl.gov/abs/1402.2577}{{\tt
  1402.2577}}].

\end{thebibliography}\endgroup
